# Crime Prediction Using Multiple-ANFIS Architecture and Spatiotemporal Data

Mashnoon Islam, Redwanul Karim, Kalyan Roy, Saif Mahmood, Sadat Hossain, Rashedur M. Rahman
Department of Electrical & Computer Engineering, North South University
Plot-15, Block-B, Bashundhara, Dhaka 1229, Bangladesh
mashnoon.islam@northsouth.edu, karim.redwanul@northsouth.edu, kalyan.roy@northsouth.edu, saif.mahmood@northsouth.edu, sadat.hossain@northsouth.edu, rashedur.rahman@northsouth.edu

*Abstract*— Statistical values alone cannot bring the whole scenario of crime occurrences in the city of Dhaka. We need a better way to use these statistical values to predict crime occurrences and make the city a safer place to live. Proper decision-making for the future is key in reducing the rate of criminal offenses in an area or a city. If the law enforcement bodies can allocate their resources efficiently for the future, the rate of crime in Dhaka can be brought down to a minimum. In this work, we have made an initiative to provide an effective tool with which law enforcement officials and detectives can predict crime occurrences ahead of time and take better decisions easily and quickly. We have used several Fuzzy Inference Systems (FIS) and Adaptive Neuro-Fuzzy Inference Systems (ANFIS) to predict the type of crime that is highly likely to occur at a certain place and time.

*Keywords*— *Fuzzy Inference System, Adaptive Neuro-Fuzzy Inference System, Fuzzification, Inference, Defuzzificaton, Root Mean Square Error (RMSE), Takagi-Sugeno Fuzzy Model, GIS, Spatiotemporal Data*

## I. Introduction

Crime has always been a major threat amongst the people in the city of Dhaka. Crime statistics is an integral source to understand the growth and declination of criminal susceptibility. In order to achieve prior information to crime, significant amount of valid and reliable data is required. The noted data by law enforcement officials is regarded as a significant source for criminal information. These data include recorded complaints filed by the victims and also by the law enforcements which ultimately gives a record of total number of criminal offences. Although each and every crime reported does not portray the actual crime situation of a definite area and sometimes some of the crime occurrences go unreported, yet police record is considered as the vital measure to perceive the crime situation. Our study focuses on the data provided by the Dhaka Metropolitan Police (DMP) on their website in order to understand the latest tendency of crime occurrences.

We have noticed a fuzzy nature in crime occurrences. Given that a crime has occurred in a particular instance of space and time, if we take the instance to be a member of a fuzzy set with a high degree of membership, the instances of space and time around that particular instance are also part of that fuzzy set, with slightly lower degrees of membership. The slightly lower degrees of membership indicate that there are still possibilities of occurrence of crime, but a bit less likely.

In this research we are using two Fuzzy Inference Systems (and consequently two Adaptive Neuro-Fuzzy Inference Systems) and the data provided by the Dhaka Metropolitan Police (DMP) on their website. The two FISs are designed for predicting murder and kidnapping respectively. We are initially conducting this research with a total amount of 349 data points, of which 184 has been used (83 for kidnapping and 101 for murder) as the dataset to demonstrate and evaluate the system. The parameters of the data are: crime type, location of the place where the crime occurred and the date of occurrence of that crime. The two FISs have been combined together to act as a single predictor. We have trained the system with the whole dataset, as the amount of data is not as high as we need it to be, and twenty percent of the data is taken as a test set. Our system requires inputs of a definite latitude, longitude and date from the user. The system automatically calculates the latitudes, longitudes and the difference between the dates that have been provided by the user with that of the nearest public holiday for every data instance. As crime occurrences such as mugging, robbery and snatching increases before and after public holidays like Eid-ul-Fitr [14], the difference (in number of days) between the enquiry date and the closest holiday for every particular crime occurrence is used as an input parameter. The system then gives a prediction of which type of crime is likely to take place in that specific location and at that specific date. The system also gives the confidence score that is associated with the prediction.

## II. Related Works

The distribution of crime is neither uniform nor random, rather it forms patterns over space and time, and is highly concentrated on just few places with high crime intensity, which are referred to as Crime Hot Spots [2]. The first law of Geography, by Tobler [3], states 'Everything is related to everything else, but near things are more related than distant things'. Hence, like every other event, crime events can also be described as interrelated events. Elaborating further, crime occurrences can be correlated by their space-time information. Malleson et al. [1] proposed crime hotspot analysis using crowd sourced data which is mainly spatial and temporal. Ratcliffe [4] reported the importance and challenges of spatiotemporal mapping to reveal the underlying patterns of crime occurrences. Hägerstrand [5] proposed a solid conceptual framework for understanding constraints on human activities such as crime. He elaborated further about the participation on crime that are influenced by those constraints, which are determined by space and time, defined as Time Geography. This framework [5] was further analyzed and formulated by Harvey J. Miller [6] for basic time geographic entities and relations. If spatiotemporal data is not analyzed using proper approaches such as proper methodology, appropriate tools, techniques, statistical tests and descriptive processes, some of the outcomes will be vague and most of the information may remain obscure to researchers. An appropriate solution can be the implication of Geographic Information Systems (GIS) [4]. Different approaches using non-hierarchical clustering, their strengths and weaknesses for predicting crime pattern analyzing spatial data are discussed in [7] by Murray et al. In [8] and [9] a fuzzy c-means clustering based method for the spatiotemporal analysis is proposed. The way that a spatial distribution of the hotspot can evolve temporally was shown in [8], but maintaining the order of the clusters was first proposed in [9] by introducing a new algorithm that preserves the order of the clusters through time, which is mentioned in the paper as Cluster Reorganization algorithm (CRA). In [10], Liu and Brown proposed a multivariate prediction model which is based upon point-patterns. These point patterns are then used to predict spatiotemporal events. These events mainly focus on the choices of the criminals for the crimes that have already taken place. Future crimes can be identified, and the hotspots of the crime can eventually be generated using these choices. In [15], Cesario, Catlett and Talia have proposed a solution where rolling time pattern of crime is predicted. The pattern gives the probing assessment executed on a two-year long prediction. An auto regression model is proposed, which is basically regression of a variable against itself. It helps achieve a significantly low average error that equals 16% for the following year crime forecast and 20% on the total of two years ahead of crime forecast. In [16], Rodríguez and Gomez proposed a solution where a spatiotemporal pattern of criminal activity is predicted. Cluster Reorganization Algorithm (CRA) is used to generate a time series and Fuzzy Clustering Method (FCM) is used for fuzzy clustering. Finally, a Memetic Algorithm, based on a combination of differential evolution and Adaptive Neuro Fuzzy Inference System (ANFIS), is used to perform forecasting on the generated time series. In [17], Baloian, Fernández and Fuentes proposed a method where crime patterns of Chilean large cities have been predicted using crime data from the Chilean Police and from freely available databases. Specific types of crime prediction has been done, such as burglaries, armed robberies and violent thefts. Prospective Method, Dempster Shafer Theory and Multi-kernel Method have been integrated together to generate a crime pattern.

Some of the other prediction tools like IBM SPSS Crime Prediction Analytics Solution (CPAS) [12] and also PredPol [11], as well as the model proposed in [10] requires complex features for analysis which are not available in the crime data of Dhaka that we are conducting our research on. Taking into consideration of this issue a probabilistic model was developed by Parvez et al. [13] to predict future street crimes in Dhaka using only spatiotemporal data, which exploits spatial and temporal proximity of past crimes for the prediction of future crimes.

III. THEORY

3.1 FUZZY INFERENCE SYSTEM

A Fuzzy Inference System uses fuzzy set theory for mapping inputs to outputs. A FIS uses "IF THEN" rules along with "AND" or "OR" connectives for creating decision-making rules. Takagi-Sugeno Fuzzy Model is one of the approaches to the Fuzzy Inference System:

*A. Fuzzification*

It is the process to convert a crisp input value into a fuzzy value with the help of membership functions. Membership functions can be of many different types such as Trapezoidal, Gaussian and Triangular.

*B. Inference*

Fuzzy inference is the process of conversion of the mapping of the inputs to an output variable with the help of the fuzzy logic.

*C. Obtaining Crsip Output Values*

The output given by the Sugeno model is generated using a linear function for each rule as in (1):

$$z_i = \alpha x_i + \beta y_i + \gamma \quad (1)$$

Finally, the weighted average of outputs of all the linear functions, as in (2), is considered to be the final output:

$$\frac{\sum_i w_i z_i}{\sum_i w_i} \quad (2)$$

Where $z_i$ is a function in p, q and r, $w_i$ is the membership degree of the rule $r_i$.

3.2 ROOT MEAN SQUARE ERROR

RMSE is a measure of how scattered the output data points are from that of the regression line.

$$RMSE = \sqrt{\frac{\sum_{i=1}^{n}(P_{prediction} - P_{output})}{n}} \quad (3)$$

Where n is the number of predicted values.

3.3 ADAPTIVE NEURO FUZZY INFERENCE SYSTEM (ANFIS)

ANFIS is basically a FIS whose parameters are adjusted using either backpropagation or a combination of backpropagation and least squares method.

## IV. DATA PRE-PROCESSING

We have collected the data from the DMP crime maps available on the website of Dhaka Metropolitan Police (DMP). The dataset contains crime occurrences of December-12, July-13 to October-13 and April-14 to June-14, making a total of around 350 data points, or in other words, around 350 data instances. A data instance of the dataset is shown in Table I:

TABLE I.  EXAMPLE DATA

| Crime Type | Address | Occurrence Date |
|---|---|---|
| Kidnapping | 7/8 Block-D, Lalmatia Girls' College Road, Dhaka Bangladesh | 30.09.13 |

We have processed the dataset to match the input parameters of our system. The example above has been converted by our data-processing Python script and the processed data is shown in Table II:

TABLE II.  PROCESSED DATA

| Crime Type | Latitude | Longitude | Day | Difference with Closest Holiday |
|---|---|---|---|---|
| kidnapping | 23.754655 | 90.369399 | 30 | 6 |

## V. DATA ANALYSIS

It is important to analyze data before designing the system. It is also important to analyze the distribution of data and also find the correlation among various data points. We have generated a scatter plot to see how crime occurrences are distributed over latitude (x-axis) and longitude (y-axis). We used different colors for each crime type. The plot for the 349 data instances is given below:

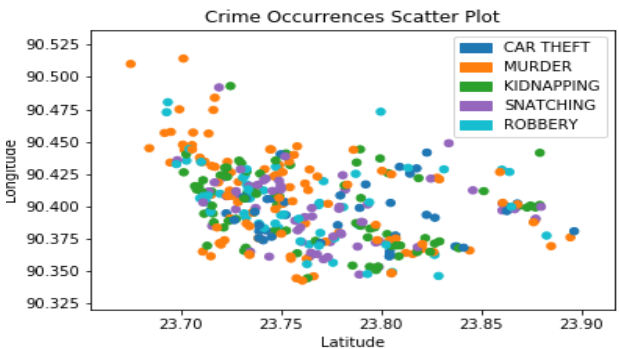

Fig. 1.  Crime Occurrences Scatter Plot

According to Fig. 1, crime occurrences are not uniformly distributed over the ranges of latitude and longitude. Also, a pattern is not clearly visible in the scatter plot. However, there is a possibility of small patterns being noticeable if we produce a magnified version of the scatter plot. This gives us a clue that using more membership functions for the input parameters "latitude" and "longitude" is going to increase the system's prediction accuracy. However, we cannot yet decide the exact number of membership functions to use just by looking at this scatter plot. To speed up the training process of our system, we have decided to use four membership functions for "latitude" and four for "longitude".

Crime rates tend to generally rise during the early and late portions of a particular month according to the data we are using, shown in Fig. 2. There is a noticeable pattern in the data we are using. This gives us a clue that using two membership functions for the input parameter "day of month" is sufficient.

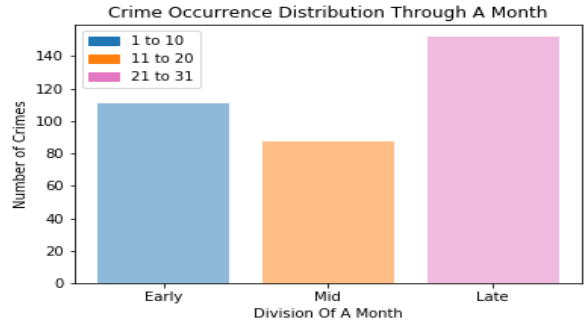

Fig. 2.  Crime Distribution Through A Month

According to our data, crime rates tend to rise before and after a public holiday, as shown in Fig.3 –Fig.5. Once again, a pattern in the data is visible. Thus, using two membership functions for the input parameter "difference with closest holiday" is sufficient.

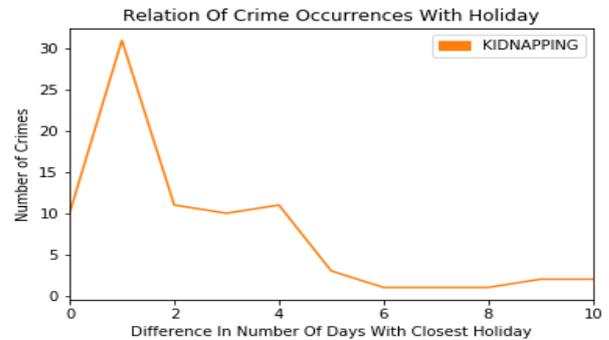

Fig. 3.  Relation Of Occurences Of Kidnapping With Holiday

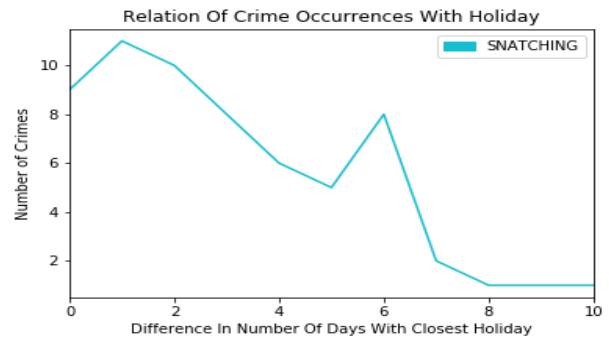

Fig. 4.  Relation Of Occurences Of Snatching With Holiday

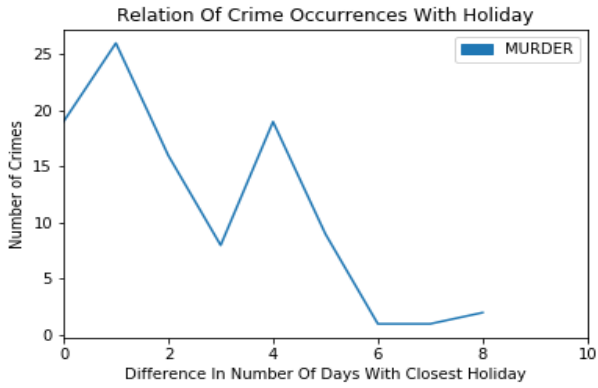

Fig. 5. Relation Of Occurences Of Murder With Holiday

As a part of our data is spatial, we are using ArcGIS to figure out the crime hotspots using Getis-Ord Gi* statistic. Incremental Spatial autocorrelation tool in ArcGIS that gives a spatial autocorrelation for a number of distances and gives a line graph of the distances and their related z-scores as output, which is shown in Fig. 6.

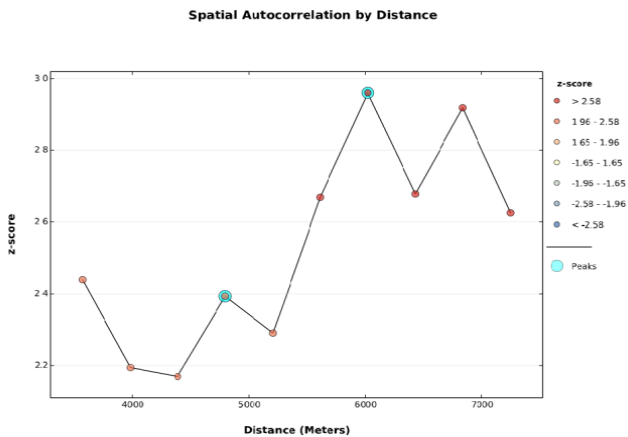

Fig. 6. Spatial Autocorrelation By Distance

The spatial clustering intensity is reflected by the z-scores. The most noticeable spatial processes that promotes clustering is based upon the statistically noticeable peak z-scores. The highest points calculated are used as input for the hotspot analyzing tool, that is, the distance radius and the z-score for each point.

After the hotspots are generated by the hotspot analyzing tool, IDW (Inverse Distance Weighted) interpolation is used to generate the heat map in Fig. 7. IDW gives a visualization of the fact that the closer the points are, the more similar they get and vice versa. The system gives the maximum weights to the points that are closest to the hotspots and the weights of points decrease with increase in the distance between the hotspots and these points, which consequently leads to the name of the method "Inverse Distance Weighted".

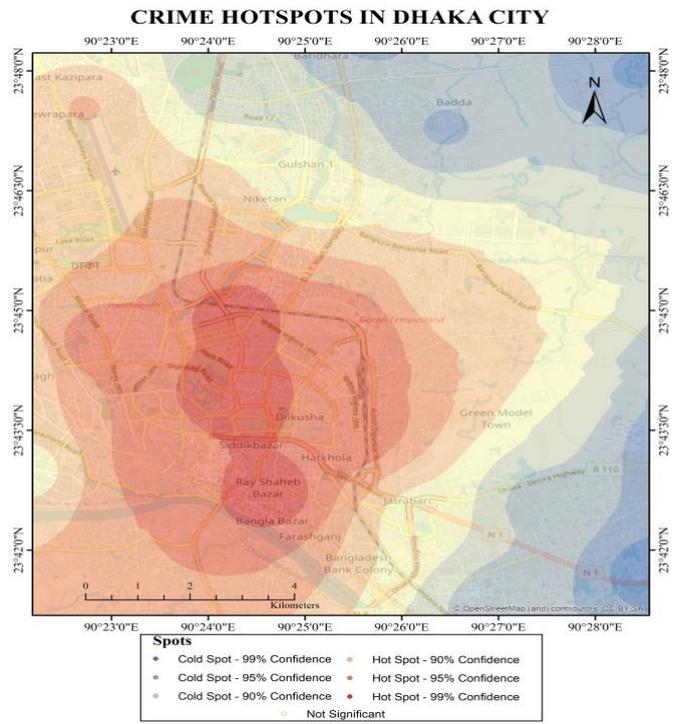

Fig. 7. Crime Heat Map Of Dhaka City

The heat map in Fig.7 is used to visualize the most crime vulnerable areas of the Dhaka City according to our data. The degrees of confidence are 90%, 95% and 99%, as shown in the heat map. Regions in this heat map are distinguished by varying shades of two colors – red and blue. In general, red regions, referred to as hot spots, are vulnerable to crime whereas blue regions, referred to as cold spots, are invulnerable to crime. The higher the confidence of a particular region, the higher the intensity of its color. For the red regions, the higher the color intensity, the higher the crime vulnerability and vice versa. For blue regions, the higher the color intensity, the higher the crime invulnerability and vice versa.

VI. METHODLOGY

A. *System Design*

The system is composed for a total of two Fuzzy Inference Systems, and each is intended to predict one type of crime. We refer to the two fuzzy inference systems as models, or in other words, "experts".

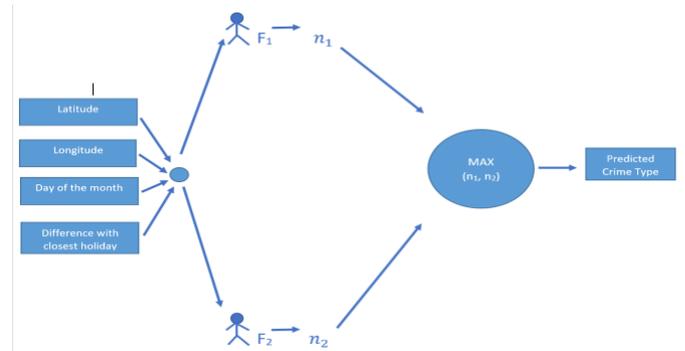

Fig. 8. Visual Representation Of The System

We can visualize the scenario of the system by looking at Fig. 8. Two crime prediction experts, $F_1$ and $F_2$, are trying to predict occurrence of the crime that they study individually. $F_1$ and $F_2$ look at a particular location, the future date that they are working on and the difference of that future date (in number of days) with the closest holiday. They individually give two different scores $n_1$ and $n_2$ respectively. The expert who gives the highest score gives the resulting prediction. That is, if $n_2$ is the highest score, then the type of crime that $F_2$ studies is the final prediction.

In our implementation, $F_1$ is the crime type "kidnapping" and $F_2$ is the crime type "murder". $n_1$ and $n_2$ are the confidence scores given by $F_1$ and $F_2$ respectively.

```
1. highest_confidence, index = 0
2. for i = 1 to number_of_fis
3.     confidence = fis[i].evaluate_fis(input_parameters)
4.     if confidence > highest_confidence
5.         highest_confidence = confidence
6.         index = i
7.     end
8. end
9. print fis[index].name
```

Fig. 9. System Pseudocode

### B. Input Parameters

The system takes four input parameters: latitude, longitude, day of the month of enquiry and difference with the holiday closest to the date of enquiry. Latitude and longitude are values that represent the position of a specific location on a geographical map.

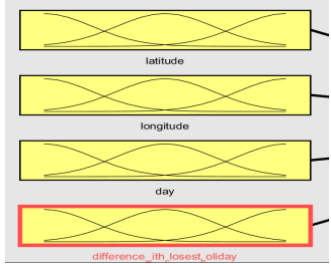

Fig. 10. Input Parameters Of The Models

### C. Output

The system gives a prediction of what it thinks the type of crime is, that will occur in the location and on the date specified by the query. Therefore, the output will be one of the labels present in the dataset.

### D. Input Membership Functions

Four Gaussian membership functions have been defined for each of the two input parameters "latitude" and "longitude" and two Gaussian membership functions have been defined for each of the two input parameters "day" and "difference with closest holiday". This is the same for each model of our system. The ranges have been set by extracting the maximum and minimum of each parameter from the processed dataset. For the functions below, the ranges have been found to be from around 23.6 to around 23.9.

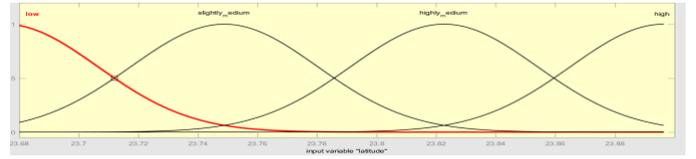

Fig. 11. Initial Input Membership Functions For Latitude

### E. Output Functions

The output functions are linear functions that give the confidence scores based on the input query. The general equation of the function is as in (3):

$$z_i = \alpha x_i + \beta y_i + \gamma \tag{3}$$

There is one output function for each rule of each model. The scoring is from 0 to 100. For now, the function parameters are assigned zero.

### F. Rule Generation

It is hard even for a detective or a policeman to get an idea about the crime occurrence situation of a place without having a look at the past statistics. Thus, we have divided the data into a uniform multi-dimensional grid, and each grid division corresponds to each of all the possible rules of each model of our system. The number of rules is 64 for each model, which is the combination of all the membership functions of each model, as shown in (4):

$$4 \times 4 \times 2 \times 2 = 64 \tag{4}$$

The criteria for a data point to be a part of a grid division is that it has to fall into the constraints that are assigned to that specific grid division. In simple words, we are finding out how much of the data supports the constraints for the grid, and consequently, how much of the data supports the rule that is associated with the grid. Let us consider three dummy data records in Table III and a two-variable fuzzy inference system with 2 membership functions, named "low" and "high" for each variable:

TABLE III. EXAMPLE DATA OF LATITUDE AND LONGITUDE

| Latitude | Longitude |
|---|---|
| 23.777 | 90.555 |
| 23.666 | 90.666 |
| 23.888 | 90.777 |

The maximum value of latitude is 23.888 and the minimum is 23.666. Thus, as shown in (5), the average of these values is:

$$average\ latitude = \left(\frac{23.666 + 23.888}{2}\right) = 23.777 \tag{5}$$

The average value of longitude is calculated in the same way and is equal to 90.666.

Therefore, one of all the possible combinations of constraints for the grid is as in (6):

$$(23.666 \leq latitude \leq 23.777)$$

$$\&$$

$$(90.555 \leq longitude \leq 90.666)$$

(6)

The constraint on the top represents the first membership function "low" for "latitude" and the constraint on the bottom represents the first membership function "low" for "longitude". The number of data points that fall into this grid division is 2, as we will see later in Table V.

All of the possible constraints of the multidimensional grid with their respective membership functions are shown in Table IV:

TABLE IV. MAPPING OF CONSTRAINTS TO MEMBERSHIP FUNCTIONS

|  | Latitude | | Longitude | |
|---|---|---|---|---|
| Constraint | 23.666 to 23.777 | 23.777 to 23.888 | 90.555 to 90.666 | 90.666 to 90.777 |
| Membership Function | low | high | low | high |

Table V shows the two-dimensional grid that the example data is divided into and the number of data points that fall into each grid division is also mentioned. Each division is represented by two constraints, and consequently two membership functions, as shown previously in Table IV:

TABLE V. THE MULTIDIMENSIONAL GRID FOR THE EXAMPLE DATA

|  |  | Longitude | |
|---|---|---|---|
|  |  | low | high |
| Latitude | low | 2 | 1 |
|  | high | 1 | 1 |

### G. Calculation Of Confidence Scores

The number of data points satisfying each grid division for each crime type is calculated. Each result is divided by the total number of data records in the dataset given, giving the probabilities of each individual grid division for each crime type. Each of the probabilities are then multiplied by 100 and set as the confidence scores for each grid. Two different data files are then generated, which contains the confidence scores for each data point. The equation is as in (7):

$$confidence \text{ of } grid = \left(\frac{total \text{ items of } grid}{total \text{ data points}}\right) \times 100$$

(7)

Taking the example from Section F, the confidence score for that grid division is as in (8):

$$confidence \text{ score of } grid = \left(\frac{2}{3}\right) \times 100 = 67\%$$

(8)

The confidence scores for all the grid divisions is given in Table VI:

TABLE VI. THE MULTIDIMENSIONAL GRID WITH CONFIDENCE SCORES

|  |  | Longitude | |
|---|---|---|---|
|  |  | low | high |
| Latitude | low | 67% | 33% |
|  | high | 33% | 33% |

Now, we have to assign each data point a confidence score generated from the process. This is straight forward – assign the confidence score of the grid division that each data point falls into. However, a data point may fall into multiple grid divisions. To solve this, we simply have taken the latest confidence score that has been calculated for a data point in our confidence score generation process. We have considered the grid divisions in the following order while calculating the scores, as in (9):

$$low - low,$$

$$low - high,$$

$$high - low,$$

$$high - high$$

(9)

As a result, we have transformed Table III by including the respective confidence scores calculated for each data point, shown in Table VII:

TABLE VII. EXAMPLE DATA WITH CONFIDENCE SCORES

| Latitude | Longitude | Confidence |
|---|---|---|
| 23.777 | 90.555 | 33% |
| 23.666 | 90.666 | 33% |
| 23.888 | 90.777 | 33% |

As previously mentioned, a data instance may fall into multiple grid divisions. For this reason, the confidence scores do not necessarily sum up to 100%.

### H. Constructing The FIS Version Of The System

We have set the output functions for the rules to simply be constant functions, as in (10):

$$z_i = \gamma_1$$

(10)

Each constant is equal to the confidence score calculated for the rule that the constant function belongs to.

### I. Constructing The ANFIS Version Of The System

We have initialized the output function parameters to be zero, as in (11):

$$z_i = 0 \times x_i + 0 \times y_i + 0$$

(11)

The two ANFIS will optimize their membership functions and linear output functions to give output equal to or very close to the training data.

*J. System Training*

The adaptive neuro-fuzzy inference systems are given the whole dataset as training data. This is because we currently have only 184 data points, which is very low for systems that use spatiotemporal data. The adaptive networks adjust the membership functions and the linear output functions to fit the training data and the confidence score for each data point in the training set is the confidence score for the grid division that the data point falls into. The adaptive networks are trained till the models' outputs and the given confidence scores in the training data are as close as possible. We can see how well the networks have been trained by calculating the root mean square errors (RMSE) of the training data and the output of the trained networks.

*K. Input Membership Functions After Training*

The input membership functions are modified by the adaptive networks to fit the training data as correctly as possible. The four input membership functions shown before have been modified by their respective adaptive network and is shown below:

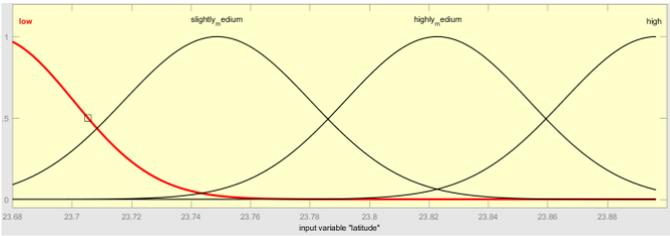

Fig. 12. Modified Input Membership Functions For Latitude

The membership function "low" has shifted slightly downwards.

*L. System Testing*

Testing the system we are proposing is different from the usual way of testing fuzzy inference systems. Since each model gives a value as output, each model alone cannot be used for classification in a convincing way. So, we give each model the input data point and get their individual outputs. The model with the highest output score is considered to be a prediction. For example, if the model trained for giving confidence scores for robbery gives the highest score, then the prediction of the system is the label "robbery".

## VII. RESULT ANALYSIS

The results confirm that crime occurrences do have a fuzzy nature, since the system could predict well even after being crafted from a very small amount of training data for each crime type that we have been granted presently, which is 83 for the crime type "kidnapping" and 101 for the crime type "murder". We have taken the last 37 data records to be the test set. The results for the FIS version of the system are shown in Table VIII:

TABLE VIII. RESULTS FOR THE FIS VERSION

| Label | Actual | Predicted | Accuracy |
|---|---|---|---|
| Kidnapping | 19 | 4 | 21% |
| Murder | 18 | 15 | 83% |
| Total | 37 | 19 | 51% |

The results for the ANFIS version of the system are in Table IX:

TABLE IX. RESULTS FOR THE ANFIS VERSION

| Label | Actual | Predicted | Accuracy |
|---|---|---|---|
| Kidnapping | 19 | 11 | 58% |
| Murder | 18 | 10 | 56% |
| Total | 37 | 21 | 57% |

Looking at the scores above, we can see that the overall prediction accuracy of the system is higher for the ANFIS version of the system. However, we can also see that the individual prediction accuracy for the crime type "murder" is lower for the ANFIS version compared to the FIS version. On the other hand, the individual prediction accuracy for the crime type "kidnapping" is lower in the FIS version compared to the ANFIS version. Thus, we decided to take the FIS version for "murder" and the ANFIS version for "kidnapping" to see whether the system produces better results. The findings are in Table X:

TABLE X. RESULTS FOR THE FIS-ANFIS HYBRID VERSION

| Label | Actual | Predicted | Accuracy |
|---|---|---|---|
| Kidnapping | 19 | 11 | 58% |
| Murder | 18 | 12 | 67% |
| Total | 37 | 23 | 62% |

The FIS-ANFIS hybrid approach has proven to be better in our test. Thus, we decided to keep this version as the final version of our system.

If given enough reliable data, we can expect the system accuracy to be above 80%, which can be validated over many datasets and situations to ensure the reliability of the method we have proposed. The increase is expected because our method has a learning architecture like ANFIS and a probabilistic approach like the one we took to calculate the confidence scores.

The results from Table VIII, Table IX, Table X are summarized in Fig. 13:

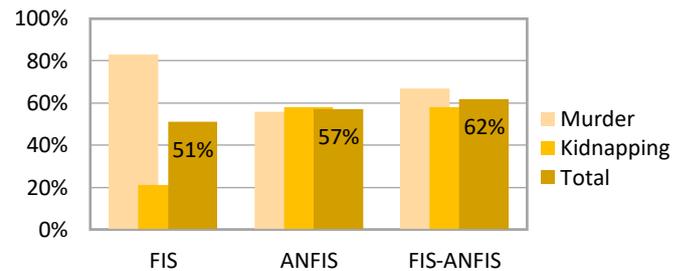

Fig. 13. Accuracy Values From Table VIII, Table IX And Table X

## VIII. Conclusion and Future Work

In this paper, we have successfully implemented a fuzzy inference model, based on previous crime occurrence data, to predict future crimes that are likely to take place. This will help the law enforcement officers and detectives prevent and take precaution against criminal incidents ahead of time.

We have shown that a fuzzy approach to crime occurrence prediction is possible. Fuzzy Inference Systems are usually used to get values as output and are not usually used for classification. However, we have shown that classification of crime occurrences using fuzzy inference systems is possible. We have also shown that the ANFIS approach does not always produce the best output, and we can take a hybrid approach to improve our system even further.

The novelty of the proposed solution is that this method has proven to work on a small dataset and does not require many input features. Big data may be, but not always, reality for the first-world countries, but not for the majority of the countries that are not blessed with Big Data tools and practices. PredPol [11] and IBM SPSS Crime Prediction Analytics Solution [12] use modern machine learning – which needs Big Data for good accuracy in prediction, and also needs higher computational power compared to the method we have proposed, which requires much less computational power. Thus, we can apply the method proposed to make a light-weight program that uses small computational power to give predictions with acceptable accuracy to be used for real life decisions and operations.

This system can be used in any region as this method finds the patterns in the data it is given, and this system can be run on machines with low computational power.

The major difficulty can be getting enough reliable multivariate data for the system. The number of input variables are probably not enough for the system to predict crime with a higher accuracy. More variables like the time of occurrence, the type of people living in an area, the peak and off-peak times of that area, the age of the majority of the people living in that area and many more may lead to a higher accurate prediction of crime. However, a lot of input features may not be required.

We are planning to use Genetic Algorithm (GA) to optimize the input parameters of the FISs and ANFISs that we used in our research. This will hopefully increase the accuracy of our system. We also aim to optimize the parameters of the ANFISs using GA, which is hopefully going to improve the overall accuracy of our system even further. We will try to build a mobile application which will predict the future crime occurrences and show a crime occurrence heat map, just like the map shown in Fig.3. The Dhaka Metropolitan Police have agreed to grant us large-scale data of crime occurrences of the last 2 years to extend our work and make this project reliable and useful to the law enforcement officers and detectives.

## IX. Acknowledgment

The crime data used for this research is provided by the Dhaka Metropolitan Police (DMP) on their website. We have taken the data from the DMP crime maps, which show crime occurrences of a particular month on the map of Dhaka and additional information related to those crime occurrences.